\documentclass[letterpaper, 10 pt, conference]{ieeeconf}

\usepackage{graphicx}
\usepackage{epstopdf}
\usepackage{epsfig}
\usepackage{times}
\usepackage{amsmath}
\usepackage{amssymb}
\usepackage{latexsym}
\usepackage{color}
\usepackage{verbatim}
\usepackage{cite}
\usepackage{BernsteinStyle2}
\usepackage[font = footnotesize]{caption}
\usepackage[font = footnotesize]{subcaption}
\usepackage[hidelinks]{hyperref}
\usepackage{diagbox}

\DeclareSymbolFont{matha}{OML}{txmi}{m}{it}
\DeclareMathSymbol{\varv}{\mathord}{matha}{118}

\usepackage{makecell}

\makeatletter
\newcommand*\bigcdot{\mathpalette\bigcdot@{.5}}
\newcommand*\bigcdot@[2]{\mathbin{\vcenter{\hbox{\scalebox{#2}{$\m@th#1\bullet$}}}}}
\makeatother

\IEEEoverridecommandlockouts

\title{\LARGE
On the Signal-to-Noise Ratio
for Numerical Differentiation%
}

\title{\LARGE
What is a Relevant Signal-to-Noise Ratio
for Numerical Differentiation?
}

\author{Shashank Verma, Mohammad Almuhaihi, and Dennis S. Bernstein%
\thanks{$^{*}$Shashank Verma, Mohammad Almuhaihi, and Dennis S. Bernstein are with the Department of Aerospace Engineering, University of Michigan, Ann Arbor, MI 48109, USA 
{\tt\small shaaero@umich.edu}}%
}

\begin{document}

\maketitle
\thispagestyle{empty}
\pagestyle{empty}

\begin{abstract}
%
%
%
In applications that involve sensor data, a useful measure of signal-to-noise ratio (SNR) is the ratio of the root-mean-squared (RMS) signal to the RMS sensor noise.
The present paper shows that, for numerical differentiation, the traditional SNR is ineffective.
In particular, it is shown that, for a harmonic signal with harmonic sensor noise, a natural and relevant SNR is given by the ratio of the RMS of the derivative of the signal to the RMS of the derivative of the sensor noise.
For a harmonic signal with white sensor noise, an effective SNR is derived.
Implications of these observations for signal processing are discussed.
\end{abstract}

\section{INTRODUCTION}

Numerical differentiation is essential for a vast range of engineering applications.
For feedback control, numerical differentiation can be used to estimate velocity from position data, thereby effectively providing a signal that might otherwise require a separate speed sensor.
For PID control, derivative feedback depends on the ability to estimate the derivative as accurately as possible and without delay \cite{PIDastrom,PID2012}.  
Flatness-based control depends on multiple derivatives \cite{Nieuwstadt1998,mboupnumeralg}. 
In control applications, performance can be degraded due to 
phase shifts and latency, which arise from filtering to reduce noise and noncausal methods that require future data. 
Numerical differentiation also plays an essential role in target tracking and trajectory prediction \cite{crouse_basic_tracking_2015}.


Numerical differentiation is an inherently difficult problem due to the fact that all real sensors are erroneous in precision and accuracy, which we collectively refer to as sensor noise.  
Sensor noise is problematic due the fact that the differentiation operator is unbounded and thus discontinuous on a dense subspace of a function space. 
The lack of continuity is manifested as amplification and thus extreme sensitivity to noise; this means that low-amplitude, high-frequency sensor noise can lead to large errors in the estimated derivative.
In addition, these errors are exacerbated by estimates of higher order derivatives.
To mitigate noise amplification, numerical differentiation techniques typically use ad hoc filters and invoke assumptions about the smoothness of the signal and properties of the sensor-noise spectrum
\cite{ramm2,Jauberteau2009,Stickel2010,zhao2013,knowles_methods,haimovich2022}.

To overcome these challenges, extensive effort has been devoted to numerical differentiation.
In a continuous-time setting, numerical differentiation has been extensively studied, with classical techniques given in 
\cite{savitzky1964smoothing,cullum,hamming}.
Modern techniques include
state-estimation methods
%
\cite{Kalata1983TheTI,tenne-2002-alpha-beta-gamma,jia_2008,karsaz_2009,lee_1999,rana_2020},
high-gain observers \cite{dabroom_discrete-time_1999}, 
deadbeat observers \cite{peng_li_non-asymptotic_2018},
polynomial approximation \cite{Liu_diff_int_2011}, 
spline-based methods \cite{ibrir_numerical_2004},
Lyapunov-function methods  \cite{polyakov_2014_diff},
and sliding-mode methods  
\cite{arie2003slidemode,reichhartinger_arbitrary-order_2018,lopez-caamal_generalised_2019,mojallizadeh2021,alwi_adap_sliding_mode_2012,LEVANT_sm_diff_1998}.

In practice, sensor data are sampled, and thus the  numerical derivative is interpreted as the sampled derivative of the underlying analog signal.
For applications involving sampled data, techniques include
\cite{Kutz2020}
as well as numerical differentiation based on adaptive input and state estimation
\cite{verma_shashank_2023_realtime_IJC,verma2024ACC}.
%


The present paper focuses on the accuracy of numerical differentiation from the perspective of the signal-to-noise ratio (SNR).
Since all real sensor data are corrupted by noise, the SNR predicts the accuracy of the estimated derivatives.
Generally speaking, a useful measure of SNR is the ratio of the root-mean-squared (RMS) signal values.
The present paper shows that, for numerical differentiation of sampled data, the traditional SNR is ineffective.
In particular, it is shown that, for a harmonic signal with harmonic sensor noise, a natural and relevant SNR is given by the ratio of the RMS of the derivative of the signal to the RMS of the derivative of the sensor noise.
For the case of white sensor noise, derivation of a relevant SNR is more subtle, and this case is considered as well.

Progress in numerical differentiation depends on the ability to assess accuracy in terms of a relevant metric.
The present paper is intended to provide relevant metrics in terms of SNR for both harmonic and white noise.
To demonstrate appropriate definitions of SNR, we present numerical differentiation results based on backward-difference differentiation as well as adaptive input and state estimation (AISE)
\cite{verma2024ACC,verma_shashank_2023_realtime_IJC}.
Additional methods for numerical differentiation can be considered as well to assess their accuracy relative to the proposed SNR's.

The contents of the paper are as follows.
Section \ref{motivation_snr} motivates the need for an introducing a new definition of SNR for numerical differentiation.
%
Section \ref{sec:num_example_harmonic} considers numerical differentiation of harmonic signals with harmonic sensor noise.
Section \ref{sec:num_example_white} focuses on numerical differentiation of harmonic signals with white sensor noise and develops a relevant SNR.
Section \ref{dis_con_fut} provides conclusions, discusses implications of this work, and proposes future work.




\section{A Motivating Example} \label{motivation_snr}

Consider the continuous-time harmonic signal and its first and second derivatives given by
\begin{align}
    y(t) &= A \sin(\omega t), \label{harmonic} \\
    \dot{y}(t) &= A \omega \cos(\omega t), \label{harmonic_sd} \\
    \ddot{y}(t) &= -A \omega^2 \sin(\omega t), \label{harmonic_dd} 
\end{align}
where $A \in \mathbb{R}$ is the amplitude and $\omega \in \mathbb{R}$ is the frequency. We assume that the measurement $y_{\rm m}(t)$ of $y(t)$ is corrupted by harmonic noise with amplitude $A_\rmn \in \mathbb{R}$ and $\omega_\rmn \in \mathbb{R}$ as
\begin{equation}
    y_{\rm m}(t) = A \sin(\omega t) + A_\rmn  \sin(\omega_\rmn  t). \label{harmonic_m_harmonic}
\end{equation}
%


For the measurement \eqref{harmonic_m_harmonic}, we define two different signal-to-noise ratios (SNR's), namely, the amplitude-based signal-to-noise ratio
\begin{align}
        \text{SNR}_{\rm amp} \isdef 20\log \frac{A}{A_\rmn}, \label{snr_amp}
\end{align}
and the more commonly used SNR
\begin{align}
        \text{SNR}_{\rm eng} \isdef 20\log \frac{{\rm RMS}(y_\rmm)}{{\rm RMS}(A_\rmn  \sin(\omega_\rmn  t))}, \label{snr_eng}
\end{align}
where ${\rm RMS}(x)$ is root-mean-squared value of $x$. 
We will investigate the relevance of \eqref{snr_amp} and \eqref{snr_eng} for numerical differentiation.

The simplest approach to differentiating a discrete-time signal is backward-difference (BD) differentiation.
As defined in \cite{PIDastrom}, let $\bfq^{-1}$ denote the backward-shift operator.
Then the {\it backward-difference single differentiator} is given by 
\begin{align}
    G_{\rm sd}(\bfq^{-1}) \isdef \cfrac{1-\bfq^{-1}}{T_\rms}, \label{sd_BD}
\end{align}
and the {\it backward-difference double differentiator} is given by
\begin{align}
    G_{\rm dd}(\bfq^{-1}) \isdef \cfrac{(1-\bfq^{-1})^2}{T_\rms^2}. \label{dd_BD}
\end{align}

The following example investigates the relevance of ${\rm SNR}_{\rm amp}$ \eqref{snr_amp} and ${\rm SNR}_{\rm eng}$ \eqref{snr_eng} for single and double numerical differentiation of a harmonic signal with harmonic sensor noise.

\begin{exam} \label{eg:harmonic_motivation}
      {\it Numerical differentiation with harmonic sensor noise.}
{\rm  
Consider the harmonic signal \eqref{harmonic} with $A = 50$ and $\omega = 1$,
%
%
which is measured in the presence of harmonic sensor noise given by
\begin{equation}
    y_\rmn(t) = \sin(\omega_\rmn t),
\end{equation}
where $\omega_\rmn \in \{0, 1, \ldots, 9, 10\}$ rad/s. The measured signal is thus given by
\begin{equation}
    y_\rmm(t) = 50 \sin(t) + \sin(\omega_\rmn t). \label{ym_exp}
\end{equation}
The signal $y_\rmm(t)$ is sampled with sample time $T_\rms = 0.01$ s. 
We compute the root-mean-squared-error (RMSE) between the single derivative estimate of \eqref{ym_exp} using BD with the true first derivative \eqref{harmonic_sd}. Similarly, for the second derivative, we compute the RMSE between the second-derivative estimate of \eqref{ym_exp} using BD with the true second derivative \eqref{harmonic_dd}.

For single differentiation of \eqref{ym_exp}, BD is used as defined by \eqref{sd_BD}.
    Figure \ref{fig:harmonicNoise_SD_moti} compares the RMSE of BD versus $\omega_\rmn$ with ${\rm SNR}_{\rm amp}$ and ${\rm SNR}_{\rm eng}$. 
    %
    It may be expected that, as the RMSE of the first-derivative estimate increases with higher values of $\omega_\rmn$, both ${\rm SNR}_{\rm amp}$ and ${\rm SNR}_{\rm eng}$ would decrease. However, Figure \ref{fig:harmonicNoise_SD_moti} shows that SNR$_{\rm amp}$ and SNR$_{\rm eng}$ are nearly constant across the range of values of $\omega_\rmn$. Therefore, ${\rm SNR}_{\rm amp}$ and ${\rm SNR}_{\rm eng}$ are not  correlated with the accuracy of the differentiation estimates.

    For double differentiation, BD is used as defined by \eqref{dd_BD}.
    Figure \ref{fig:harmonicNoise_DD_moti} compares the RMSE of BD versus $\omega_\rmn$ with ${\rm SNR}_{\rm amp}$ and ${\rm SNR}_{\rm eng}$. 
    %
    %
    As for the first derivative, ${\rm SNR}_{\rm amp}$ and ${\rm SNR}_{\rm eng}$ for the second derivative are not correlated with the differentiation accuracy.

To address these deficiencies, Sections \ref{sec:num_example_harmonic} and \ref{sec:num_example_white} present new definitions of SNR that are more correlated with the accuracy of the derivative estimates.
%


    %

 \begin{figure}[h!t]
              \begin{center}{\includegraphics[width=1.0\linewidth]{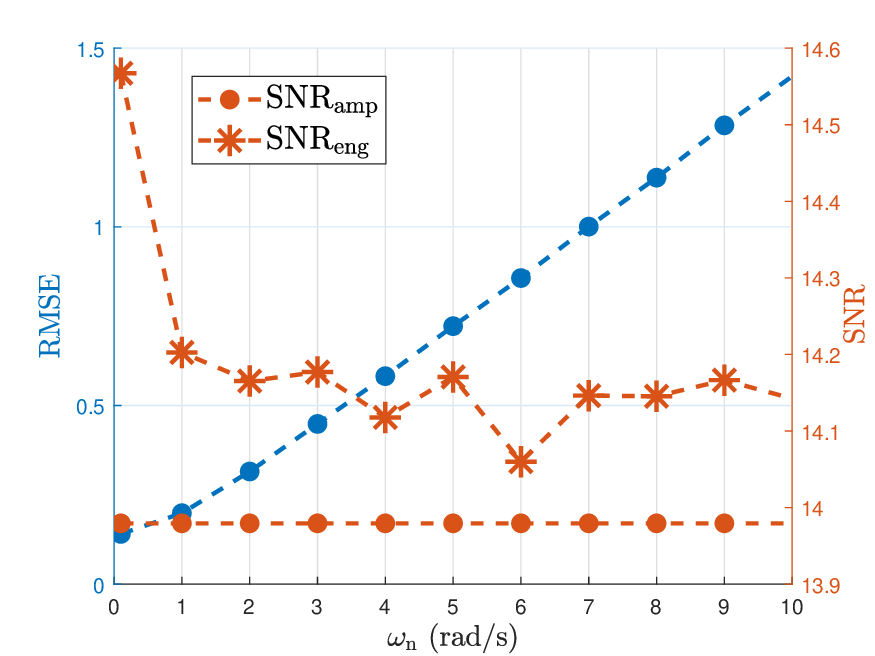}}
            \end{center}
            \caption{ {\it  Example \ref{eg:harmonic_motivation}: Numerical differentiation with harmonic sensor noise using BD.} The left axis label shows the RMSE of the estimate of a single derivative of $ y_{\rm m}$ versus $\omega_{\rmn}$. The right axis label shows ${\rm SNR}_{\rm amp}$ and ${\rm SNR}_{\rm eng}$.  Note that neither SNR correlates with the accuracy of the estimates.} 
        \label{fig:harmonicNoise_SD_moti}
\end{figure}
%
%
 \begin{figure}[h!t]
              \begin{center}{\includegraphics[width=1.0\linewidth]{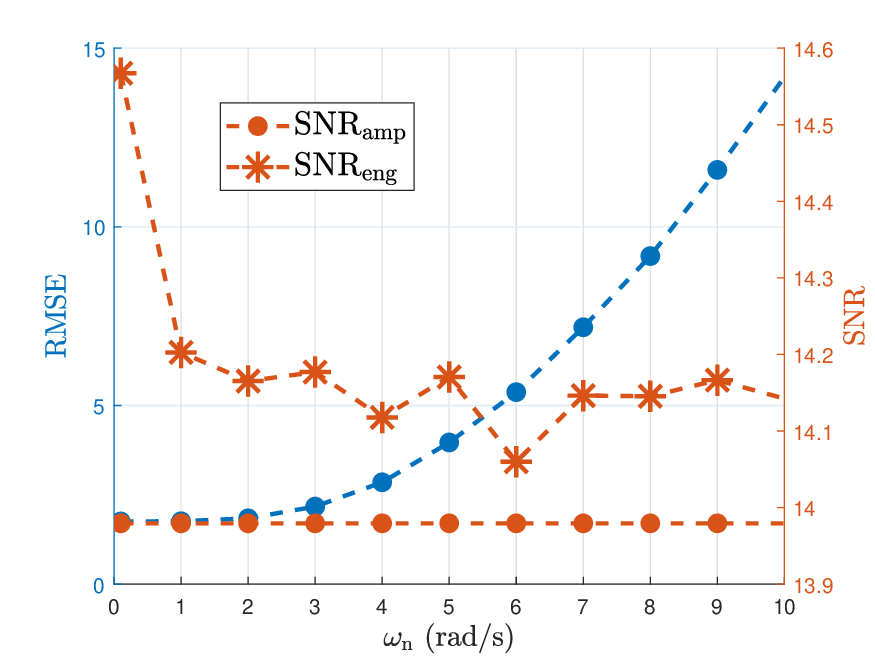}}
            \end{center}
            \caption{ {\it  Example \ref{eg:harmonic_motivation}: Numerical differentiation with harmonic sensor noise using BD.} The left axis label shows the RMSE of the estimate of a double derivative of $y_{\rm m}$ versus $\omega_{\rmn}$. The right axis label shows ${\rm SNR}_{\rm amp}$ and ${\rm SNR}_{\rm eng}$.  Note that neither SNR correlates with the accuracy of the estimates.} 
        \label{fig:harmonicNoise_DD_moti}
\end{figure}

}
\end{exam}






\section{Numerical Differentiation of Harmonic Signals with Harmonic Noise}  \label{sec:num_example_harmonic}
%



%
%
The first and second derivatives of \eqref{harmonic_m_harmonic} are given by
\begin{align}
    \dot y_{\rm m}(t) &= A \omega \cos(\omega t) + A_\rmn  \omega_\rmn  \cos(\omega_\rmn  t), \label{harmonic_m_sd} \\
    \ddot y_{\rm m}(t) &= -A \omega^2 \sin(\omega t) - A_\rmn  \omega_\rmn ^2 \sin(\omega_\rmn  t), \label{harmonic_m_dd}
\end{align}
For convenience, we define the single-differentiation error 
\begin{align}
    e_{\rm sd}(t) &\isdef \dot y_{\rm m}(t) - \dot y(t)\nn\\  &=  A_\rmn  \omega_\rmn  \cos(\omega_\rmn t),
\end{align}
and the double-differentiation error 
\begin{align}
    e_{\rm dd}(t) &\isdef \ddot y_{\rm m}(t) - \ddot y(t)\nn\\  &=  -A_\rmn  \omega_\rmn^2  \sin(\omega_\rmn t).
\end{align}
The exact root-mean-squared error (RMSE) over one period $T = \frac{2\pi}{\omega}$ for single and double differentiation are thus given by
\begin{align}
        \text{RMSE}_\text{sd} 
        &= \sqrt{\frac{1}{T} \int_0^T e_{\rm sd}(t)^2\, \rmd t} \nn\\
        &= A_\rmn  \omega_\rmn  \sqrt{\frac{1}{T} \int_0^T \cos^2(\omega_\rmn  t) \, \rmd t} 
        \nn\\
        &=\frac{A_\rmn  \omega_\rmn }{\sqrt{2}}. \label{sd_rmse_harmonic}
\end{align}
Similarly, for the second derivative,
\begin{align}
        \text{RMSE}_\text{dd} 
        &= \sqrt{\frac{1}{T} \int_0^T e_{\rm dd}(t)^2 \, \rmd t}\nn \\
        &= A_\rmn  \omega_\rmn ^2 \sqrt{\frac{1}{T} \int_0^T \sin^2(\omega_\rmn  t) \, \rmd t} \nn \\
        &=\frac{A_\rmn  \omega_\rmn ^2}{\sqrt{2}}.\label{dd_rmse_harmonic}
\end{align}
We thus define the single-differentiation and double-differentiation signal-to-noise ratios by
\begin{align}
        \text{SNR}_{\rm sd} \isdef \frac{1}{A_\rmn  \omega_\rmn },
         \\
        \text{SNR}_{\rm dd} \isdef \frac{1}{A_\rmn  \omega_\rmn ^2}.
\end{align}
Therefore,
\begin{align}
    \text{RMSE}_{\rm sd}  = \frac{1}{ \sqrt{2}  \, \text{SNR}_{\rm sd}}, \\
    \text{RMSE}_{\rm dd}  = \frac{1}{\sqrt{2} \, \text{SNR}_{\rm dd}}.
\end{align}
%
%

    
    In addition to BD differentiation, we consider adaptive input and state estimation (AISE) 
    \cite{ verma_shashank_2023_realtime_IJC,verma2024ACC} for real-time numerical differentiation to estimate first and second derivatives of $y_\rmm(t)$ in the following examples.

\begin{exam} \label{eg:nd_harmonic_noise}
      {\it Numerical differentiation with harmonic sensor noise.}
{\rm To illustrate  ${\rm RMSE}_{\rm sd}$ and ${\rm RMSE}_{\rm dd}$ as a function of the harmonic noise frequency $\omega_\rmn$, as given in \eqref{sd_rmse_harmonic} and \eqref{dd_rmse_harmonic}, we consider \eqref{harmonic} with  $A = 1$ and  $\omega = 2\pi$ rad/s. For harmonic noise, $A_\rmn = 0.2$ in \eqref{harmonic_m_harmonic}. The signal $y_{\rm m}(t)$ is sampled with sampling period $T_\rms = 0.01$ s.

For single differentiation using AISE, let $n_\rme = 25$, $n_\rmf = 50$, $R_z = 1$, $R_d = 10^{-6}$, $R_\theta = 10^{-3}I_{51}$, $\eta = 0.002, \tau_n = 5, \tau_d = 25, \alpha = 0.002$, and $R_{\infty} = 10^{-4}.$ 
    The parameters $V_{1,k}$ and $V_{2,k}$ are adapted, with $\eta_{L} = 10^{-6}$, $\eta_{\rmU} = 0.1$, and $\beta = 0.5$. 
    For BD, we use \eqref{sd_BD}.
    %
    %
    Figure \ref{fig:HarmonicNoise_SD_SNR} shows that ${\rm SNR}_{\rm sd}$ in orange (right axis label) decreases as the ${\rm RMSE}_{\rm sd}$ in blue (left axis label) increases with increasing $\omega_\rmn$. The ${\rm RMSE}_{\rm sd}$ values for single differentiation using BD and AISE follow the exact RMSE \eqref{sd_rmse_harmonic}.

For double differentiation using AISE, let $n_\rme = 25$, $n_\rmf = 50$, $R_z = 1$, $R_d = 10^{-7}$, $R_\theta = 10^{-6}I_{51}$, $\eta = 0.002, \tau_n = 5, \tau_d = 25, \alpha = 0.002$, and $R_{\infty} = 10^{-4}.$ 
    The parameters $V_{1,k}$ and $V_{2,k}$ are adapted, with $\eta_{L} = 10^{-6}$, $\eta_{\rmU} = 0.1$, and $\beta = 0.5$. 
    For BD, we use \eqref{dd_BD}.
    %
    %
     %
     Figure \ref{fig:HarmonicNoise_DD_SNR} shows that ${\rm SNR}_{\rm dd}$ in orange (right axis label) decreases as the ${\rm RMSE}_{\rm dd}$ in blue (left axis label) increases with increasing $\omega_\rmn$. The ${\rm RMSE}_{\rm dd}$ values for double differentiation using BD and AISE follow the exact RMSE \eqref{dd_rmse_harmonic}.


\begin{figure}[h!t]
              \begin{center}{\includegraphics[width=1.0\linewidth]{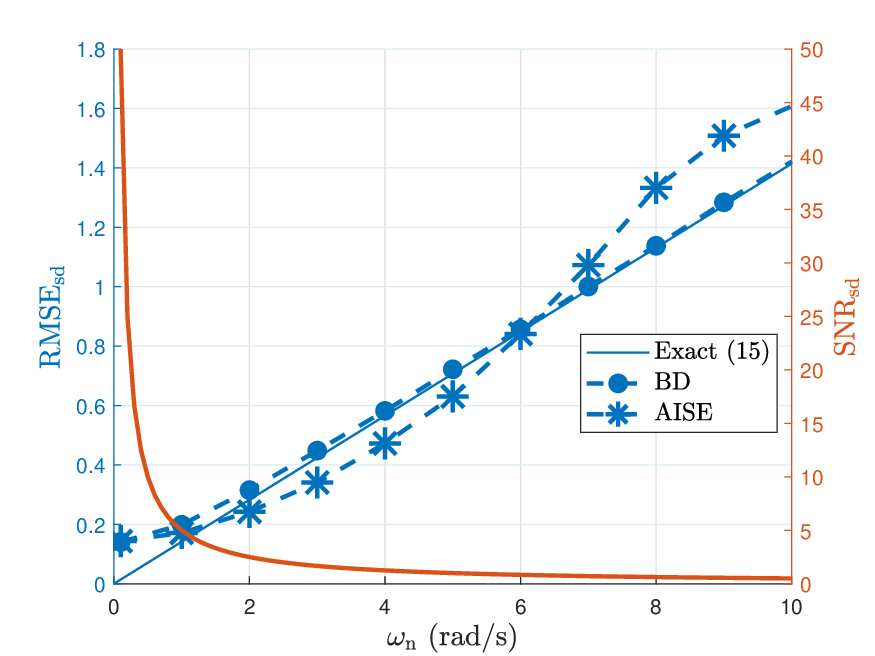}}
            \end{center}
            \caption{ {\it  Example \ref{eg:nd_harmonic_noise}: Numerical differentiation with harmonic sensor noise.} The left axis label shows the ${\rm RMSE}_{\rm sd}$ of the estimates of $\dot y_{\rm m}$ versus $\omega_\rmn$. The derivative is computed using BD, AISE, and exact \eqref{sd_rmse_harmonic}. The right axis label shows ${\rm SNR}_{\rm sd}$.  Note that SNR$_{\rm sd}$ is highly correlated with the accuracy of the estimates.
            } 
        \label{fig:HarmonicNoise_SD_SNR}
\end{figure}
 \begin{figure}[h!t]
              \begin{center}{\includegraphics[width=1.0\linewidth]{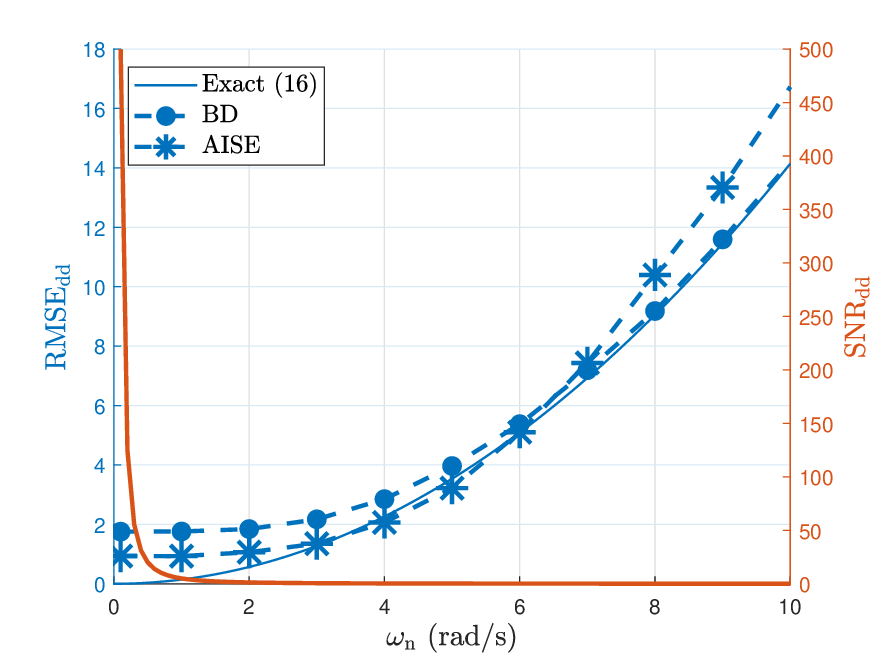}}
            \end{center}
            \caption{ {\it  Example \ref{eg:nd_harmonic_noise}: Numerical differentiation with harmonic sensor noise.} The left axis label shows the ${\rm RMSE}_{\rm dd}$ of the estimates of $\ddot y_{\rm m}$ versus $\omega_\rmn$. The derivative is computed using BD, AISE, and exact \eqref{dd_rmse_harmonic}. The right axis label shows ${\rm SNR}_{\rm dd}$. Note that SNR$_{\rm dd}$ is highly correlated with the accuracy of the estimates.
            } 
        \label{fig:HarmonicNoise_DD_SNR}
\end{figure}

}
\end{exam}

\section{Numerical Differentiation of Harmonic Signals with White Noise}  \label{sec:num_example_white}

We now consider harmonic signals corrupted by white noise.  
Hence, let 
\begin{align}
    y_{\rm m}(t) = A \sin(\omega t) + \eta(t),
\end{align}
where  $\eta \sim \mathcal{N}(0,\sigma).$
%
%
The signal-to-noise ratio (SNR) of the $y_{\rm m}(t)$ is defined as 
\begin{align}
        \text{SNR}_{0} \isdef \frac{A}{\sigma}. \label{snr_0_W}
\end{align}
Formally, we write
\begin{align}
    \dot y_{\rm m}(t) &= A \omega \cos(\omega t) + \dot \eta(t),\\
    \ddot y_{\rm m}(t) &= -A \omega^2 \sin(\omega t) + \ddot \eta(t).
\end{align}
For convenience, we define the single-differentiation error 
\begin{align}
    e_{\rm sd}(t) &\isdef \dot y_{\rm m}(t) - \dot y(t)\nn\\  &=  \dot \eta(t),
\end{align}
and the double-differentiation error 
\begin{align} 
    e_{\rm dd}(t) &\isdef \ddot y_{\rm m}(t) - \ddot y(t)\nn\\  &=  \ddot \eta(t).
\end{align}

Since $\eta$ is not differentiable, we focus on the sampled values of $\eta.$
In particular, we consider the approximation for single-differentiation as
\begin{equation}
    \dot \eta_k \approx \frac{\eta_k - \eta_{k-1}}{T_\rms} \label{eta_approx_discrete_sd}
\end{equation}
where $k \geq 0$ is the step, $\eta_k \isdef \eta (k T_\rms)$, and $T_\rms$ is the sample time.
Using \eqref{eta_approx_discrete_sd}, we define the random variable  
\begin{align}
    \zeta \isdef \frac{\eta_k - \eta_{k-1}}{T_\rms}. 
\end{align}
Since samples of white noise $\eta_k$ are uncorrelated, the variance of $\zeta$ is given by
\begin{align}
    \text{Var}(\zeta) &= \frac{\text{Var}(\eta_k) +\text{Var}(\eta_{k-1}) }{T_\rms^2} = \frac{2\sigma^2}{T_\rms^2}.
\end{align}
Hence,
\begin{align}
        \zeta \sim \mathcal{N}\bigg(0, \frac{\sqrt{2}\sigma}{T_\rms}\bigg). 
\end{align}

%
%
Similarly, for double differentiation,
\begin{equation}
    \ddot \eta_k \approx \frac{\eta_k - 2\eta_{k-1} + \eta_{k-2}}{T_\rms^2}. \label{eta_approx_discrete_dd}
\end{equation}
Using \eqref{eta_approx_discrete_dd}, we define the random variable  
\begin{align}
    \beta \isdef \frac{\eta_k - 2\eta_{k-1} + \eta_{k-2}}{T_\rms^2},
\end{align}
whose variance is
\begin{align}
        \text{Var}(\beta) &= \frac{\text{Var}(\eta_k) + 4\text{Var}(\eta_{k-1}) +\text{Var}(\eta_{k-2}) }{T_\rms^4} = \frac{6\sigma^2}{T_\rms^4}.
        \label{eq: White-DDRMS}
\end{align}
Hence, 
\begin{align}
    \beta \sim \mathcal{N}(0, \frac{\sqrt{6}\sigma}{T_\rms^2}).
\end{align}
For single and double differentiation with white sensor noise, the RMSE values are thus given by
\begin{align}
\text{RMSE}_\text{sd} = \frac{\sqrt{2}\sigma}{T_\rms}, \label{sd_rmse_white} \\
    \text{RMSE}_\text{dd} = \frac{\sqrt{6}\sigma}{T_\rms^2}.  \label{dd_rmse_white}
\end{align}
We thus define 
\begin{align}
    \text{SNR}_\text{sd} \isdef \frac{T_\rms}{\sigma}, \label{prop_snr_white_sd}\\
    \text{SNR}_\text{dd} \isdef \frac{T_\rms^2}{\sigma}. \label{prop_snr_white_dd}
\end{align}
Therefore,
\begin{align}
    \text{RMSE}_\text{sd}  = \frac{1}{\sqrt{2} \, \text{SNR}_\text{sd}}, \\
     \text{RMSE}_\text{dd} = \frac{1}{\sqrt{6} \,  \text{SNR}_\text{dd}}.
\end{align}
Furthermore, it follows from \eqref{sd_rmse_white} and \eqref{dd_rmse_white} that
\begin{align}
        \text{RMSE}_\text{dd} &= \frac{\sqrt{6}\sigma}{T_{\rm s}^2} =  \frac{\sqrt{6}}{T_{\rm s} \sqrt{2}} \left( \frac{\sqrt{2}}{T_{\rm s}} \sigma \right)  
        = \frac{\sqrt{3}}{T_{\rm s}}\text{RMSE}_\text{sd},
    \label{eq:white-higher-orders}
\end{align}
which implies
\begin{align}
    \frac{\rm RMSE_{\rm dd}}{\rm RMSE_{\rm sd}} = \frac{\sqrt{3}}{T_\rms}. \label{rmse_ratio}
\end{align}
%
%


%

\begin{exam} \label{eg:nd_white_noise}
      {\it Numerical differentiation with white sensor noise.}
{\rm To illustrate  ${\rm RMSE}_{\rm sd}$ and ${\rm RMSE}_{\rm dd}$ as a function of the white noise, as given in \eqref{sd_rmse_white} and \eqref{dd_rmse_white}, we consider \eqref{harmonic} with  $A = 1$ and  $\omega = 2\pi$ rad/s. For white noise, $\sigma = 1$. The signal $y_{\rm m}(t)$ is sampled with sampling period $T_\rms = 0.01$ s.

For Single Differentiation using AISE, let $n_\rme = 25$, $n_\rmf = 50$, $R_z = 1$, $R_d = 10^{-6}$, $R_\theta = 10^{-3}I_{51}$, $\eta = 0.002, \tau_n = 5, \tau_d = 25, \alpha = 0.002$, and $R_{\infty} = 10^{-4}.$ 
    The parameters $V_{1,k}$ and $V_{2,k}$ are adapted, with $\eta_{L} = 10^{-6}$, $\eta_{\rmU} = 0.1$, and $\beta = 0.55$. 
    For BD, we use \eqref{sd_BD}.
    %
    %
    Figure \ref{fig:whiteNoise_SD} shows that ${\rm SNR}_{\rm sd}$ in orange (right axis label) decreases as the ${\rm RMSE}_{\rm sd}$ in blue (left axis label) increases. The ${\rm RMSE}_{\rm sd}$ values for single differentiation using BD and AISE follow the exact RMSE \eqref{sd_rmse_white}, where AISE has lower RMSE due to the adaptive filtering provided by AISE.

For double differentiation using AISE, let $n_\rme = 25$, $n_\rmf = 50$, $R_z = 1$, $R_d = 10^{-7}$, $R_\theta = 10^{-6}I_{51}$, $\eta = 0.002, \tau_n = 5, \tau_d = 25, \alpha = 0.002$, and $R_{\infty} = 10^{-4}.$ 
    The parameters $V_{1,k}$ and $V_{2,k}$ are adapted, with $\eta_{L} = 10^{-6}$, $\eta_{\rmU} = 0.1$, and $\beta = 0.55$. 
    For BD, we use \eqref{dd_BD}.
    %
    %
    %
    Figure \ref{fig:whiteNoise_DD} shows that ${\rm SNR}_{\rm dd}$ in orange (right axis label) decreases as the ${\rm RMSE}_{\rm dd}$ in blue (left axis label) increases. The ${\rm RMSE}_{\rm dd}$ values for double differentiation using BD and AISE follow the exact RMSE \eqref{dd_rmse_white}, where AISE has lower RMSE due to the adaptive filtering provided by AISE.
    Figure \ref{fig:whiteNoise_case_ratio} empirically validates the ratio of RMSE of the second derivative and the first derivative of $y_{\rm m}$ versus $\sigma_{\rm noise}$ \eqref{rmse_ratio}. In particular, \eqref{rmse_ratio} computed using BD and  AISE follow the trend of the theoretical ratio $\frac{\sqrt{3}}{T_\rms}$ of \eqref{rmse_ratio}.


\begin{figure}[h!t]
              \begin{center}{\includegraphics[width=1.0\linewidth]{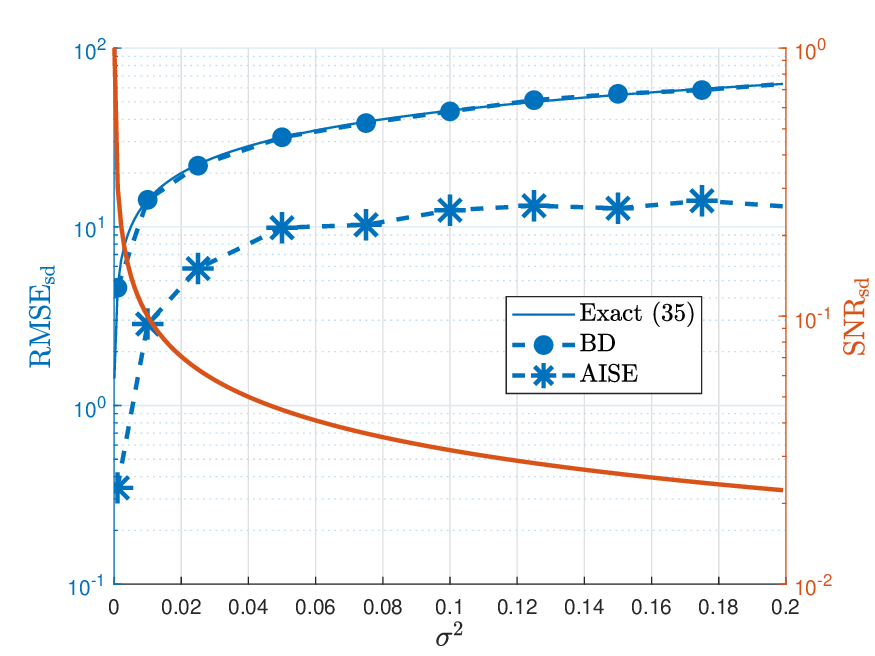}}
            \end{center}
            \caption{ {\it  Example \ref{eg:nd_white_noise}: Numerical differentiation with white sensor noise.} The left axis label shows ${\rm RMSE}_{\rm sd}$ of $\dot y_{\rm m}$ versus $\sigma^2$. The derivatives are computed using BD, AISE, and exact \eqref{sd_rmse_white}. The right axis label shows ${\rm SNR}_{\rm sd}$ \eqref{prop_snr_white_sd}.
            Note that SNR$_{\rm sd}$ is highly correlated with the accuracy of the estimates.
            } 
        \label{fig:whiteNoise_SD}
\end{figure}
 \begin{figure}[h!t]
              \begin{center}{\includegraphics[width=1.0\linewidth]{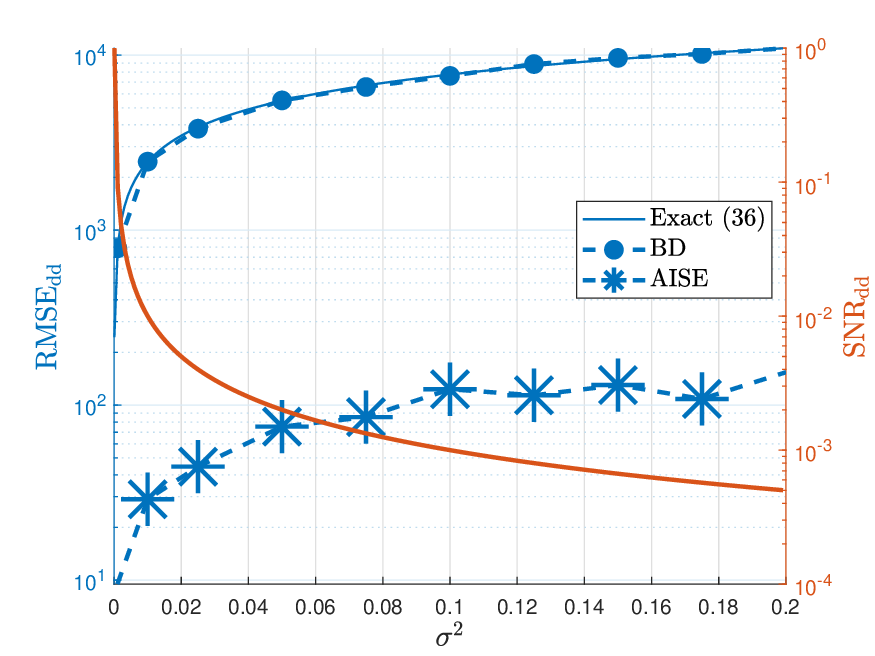}}
            \end{center}
            \caption{ {\it  Example \ref{eg:nd_white_noise}: Numerical differentiation with white sensor noise.} The left axis label shows ${\rm RMSE}_{\rm dd}$ of $\ddot y_{\rm m}$ versus $\sigma^2$. The derivatives are computed using BD, AISE, and exact \eqref{dd_rmse_white}. The right axis label shows ${\rm SNR}_{\rm dd}$ \eqref{prop_snr_white_dd}.
            Note that SNR$_{\rm dd}$ is highly correlated with the accuracy of the estimates.
            } 
        \label{fig:whiteNoise_DD}
\end{figure}

 \begin{figure}[h!t]
              \begin{center}{\includegraphics[width=1.0\linewidth]{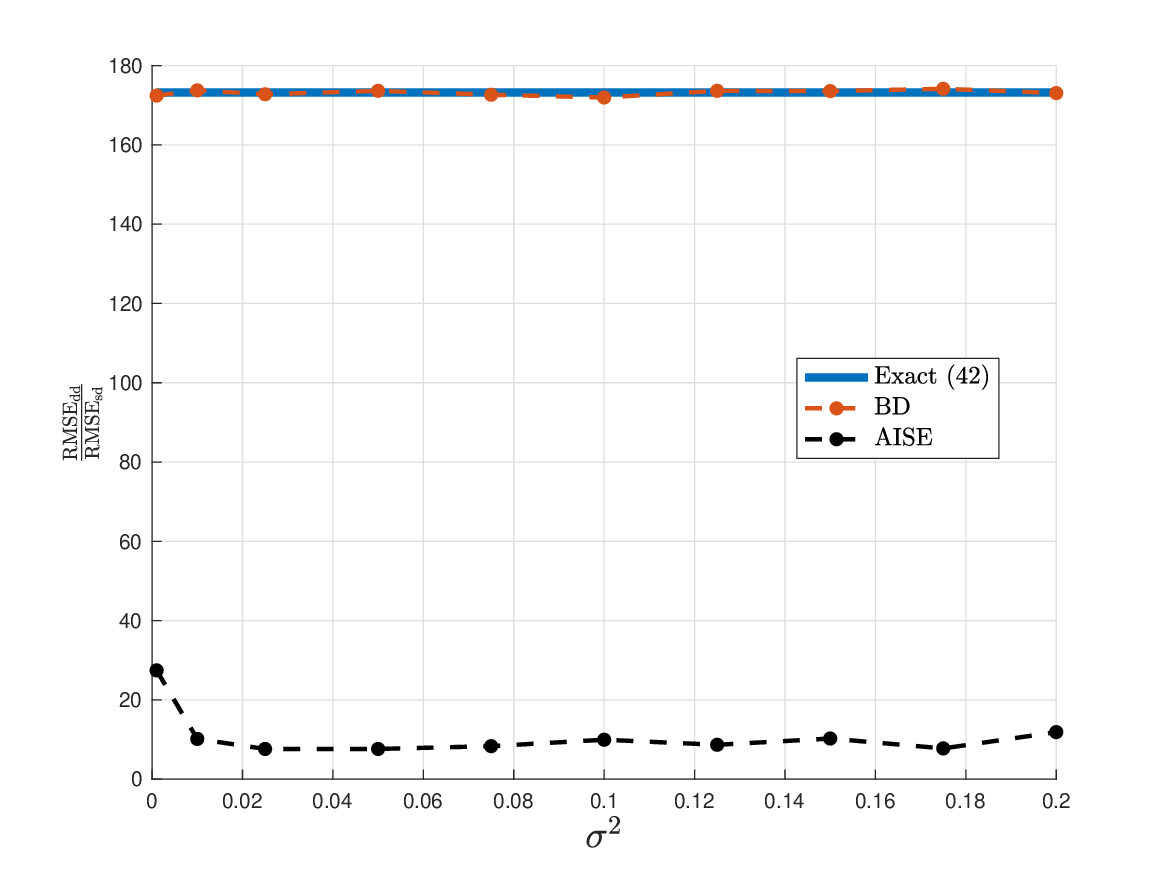}}
            \end{center}
            \caption{{\it  Example \ref{eg:nd_white_noise}: Numerical differentiation with white sensor noise.} 
            %
            %
            Ratio of RMSE of the second derivative to the RMSE of the first derivative of $y_{\rm m}$ versus $\sigma_{\rm noise}$.
            The closeness of the exact ratio and the BD estimates confirms \eqref{rmse_ratio}.
            Note that AISE is more accurate than BD due to the adaptive filtering.} 
        \label{fig:whiteNoise_case_ratio}
\end{figure}

}
\end{exam}

\section{DISCUSSION AND FUTURE RESEARCH} \label{dis_con_fut}

For applications that depend on sensor data, a relevant SNR is essential for predicting the accuracy of data-based estimates and thus for determining sensor requirements.
For numerical differentiation it is perhaps not surprising that a relevant SNR is not the ratio of the RMS of signal to the RMS of the sensor noise, but rather is the ratio of the RMS of the derivative of the signal to the RMS of the derivative of the sensor noise.
For harmonic signals with harmonic sensor noise, it is straightforward to propose a relevant SNR;  it is less obvious how to do this for white noise, especially in the practically relevant case of sampled data.

The main contributions of this paper were thus 1) motivation for the need for relevant SNR's for numerical differentiation, and 2) derivation of expressions for these SNR's for harmonic and white sensor noise. 
These SNR's were illustrated using backward difference (BD) and adaptive input and state estimation (AISE).
Similar numerical investigations can be performed for alternative numerical differentiation algorithms.
In particular, the Kalman filter based on integrator dynamics
is a convenient approach to numerical differentiation
\cite{Kalata1983TheTI,tenne-2002-alpha-beta-gamma,jia_2008,karsaz_2009,lee_1999,rana_2020},
and we expect that the SNR's proposed in the present paper are relevant for these techniques.
The broader implications of this paper for the Kalman filter are that a relevant SNR cannot be naively defined based solely on the RMS of the signal and the sensor noise.
Future research will explore this issue.


\section*{ACKNOWLEDGMENTS}
This research supported by NSF grant CMMI 2031333 and SACM.
The authors are grateful for helpful discussions with David Crouse.

\bibliography{bibpaper,bibijc}
\bibliographystyle{ieeetr}

\end{document}